\documentclass[journal]{IEEEtran}
\usepackage{cite}
\usepackage{xspace,amsmath,amssymb,amsfonts,epsfig,subfigure,syntonly}
\usepackage{cite,bm,color,url,textcomp,empheq,boxedminipage}
\usepackage{algorithmicx,algorithm}
\usepackage{epstopdf,makecell}
\usepackage{empheq}
\usepackage{pifont}
\usepackage{soul}
\usepackage[numbers,sort&compress]{natbib}
\usepackage{graphicx,graphics}  
\usepackage{multirow,multicol}
\usepackage{psfrag}    
\usepackage{stfloats}
\usepackage{url}

\newcommand{\mv}[1]{\mbox{\boldmath{$ #1 $}}}

\long\def\symbolfootnote[#1]#2{\begingroup
\def\thefootnote{\fnsymbol{footnote}}
\footnote[#1]{#2}\endgroup}

\psfull

\allowdisplaybreaks[4]

\begin{document}
\title{Placement Optimization for UAV-Enabled Wireless Networks with Multi-Hop Backhauls}


\author{Peiming~Li and Jie~Xu\textsuperscript{(*)}\\
\thanks{Part of this paper has been submitted to the IEEE International Conference on Communication Systems (ICCS), Chengdu, China, Dec. 19--21, 2018.}
\thanks{This work was supported in part by the National Natural Science Foundation of China (Project No. 61871137).}
\thanks{The authors are with the School of Information Engineering, Guangdong University of Technology, Guangzhou, 510006, China (e-mail: peiminglee@outlook.com, jiexu@gdut.edu.cn). J. Xu is the corresponding author.}

}

\maketitle
\begin{abstract}
Unmanned aerial vehicles (UAVs) have emerged as a promising solution to provide wireless data access for ground users in various applications (e.g., in emergency situations). This paper considers a UAV-enabled wireless network, in which multiple UAVs are deployed as aerial base stations (BSs) to serve users distributed on the ground. Different from prior works that ignore UAVs' backhaul connections, we practically consider that these UAVs are connected to the core network through a ground gateway node via rate-limited multi-hop wireless backhauls. We also consider that the air-to-ground (A2G) access links from UAVs to users and the air-to-air (A2A) backhaul links among UAVs are operated over orthogonal frequency bands. Under this setup, we aim to maximize the common (or minimum) throughput among all the ground users in the downlink of this network subject to the flow conservation constraints at the UAVs, by optimizing the UAVs' deployment locations, jointly with the bandwidth and power allocation of both the access and backhaul links. However, the common throughput maximization is a non-convex optimization problem that is difficult to be solved optimally. To tackle this issue, we use the techniques of alternating optimization and successive convex programming (SCP) to obtain a locally optimal solution. Numerical results show that the proposed design significantly improves the common throughput among all ground users as compared to other benchmark schemes.
\end{abstract}



\section{Introduction}\label{sec:introduction}
\IEEEPARstart{U}{nmanned} aerial vehicles (UAVs), also known as (a.k.a.) drones, have found a wide range of applications in, e.g., cargo delivery, aerial inspection, precision agriculture, and traffic monitoring. Among others, employing UAVs as aerial communication platforms to assist terrestrial wireless communications has recently emerged as one of the key technologies for the fifth-generation (5G) cellular networks, which has attracted a lot of interests from both academia and industry \cite{ZYwireless,SAR}. For example, UAVs can be used as aerial base stations (BSs) to provide basic wireless data access for remote areas and in emergency situations (e.g., after natural disasters), as well as to enhance the network capacity in terrestrial hot spots \cite{J.Lyu, CZ, MM, 2D3D2, optimal}. On the other hand, UAVs can be utilized as aerial relays to help far-apart ground users exchange information \cite{zeng2016throughput, radio1}, and as access points in the sky for information dissemination and data collection with ground nodes (e.g., sensors and actuators in Internet-of-things (IoT) networks) \cite{C.Zhan2}. Besides wireless communications, UAVs can also be used as aerial platforms for wireless power transfer (WPT) \cite{WPT1, JieXuWPT}, wireless powered communication networks (WPCN) \cite{LX}, and mobile edge computing (MEC) \cite{MEC}. In the industry, various companies have launched their UAV-assisted wireless communication projects, and some preliminary prototypes include Facebook's Aquila \cite{fb} and Nokia's flying-cell (F-Cell) \cite{nokia}.

As compared to conventional terrestrial wireless communications, UAV-assisted wireless communications have the following advantages. First, UAV-enabled aerial communication platforms can be quickly deployed on demand, and thus are cost-effective and suitable for emergency scenarios, e.g., when the terrestrial wireless infrastructures are damaged due to natural disasters. Next, the air-to-ground (A2G) wireless channels between UAVs and ground nodes normally have much stronger line-of-sight (LoS) links than conventional ground-to-ground (G2G) wireless channels; as a result, the aerial BSs are expected to provide better wireless coverage and higher communication throughput than ground BSs. Furthermore, due to the fully controllable mobility in three-dimensional (3D) airspace, UAVs can adaptively change their locations over time for reducing the distances with intended ground users, so as to further improve the communication performance.

In general, UAV-enabled aerial wireless communication platforms can be classified into two categories depending on whether UAVs are quasi-stationary or fully mobile. For quasi-stationary UAVs, prior works \cite{2D3D1, 2D3D2, optimal, MM, CZ, radio1, radio2} focused on the optimization of their deployment locations for communication performance improvement. For instance, \cite{2D3D2} and \cite{2D3D1} optimized the two-dimensional (2D)/3D placement of UAVs to maximize the network revenue (i.e., the number of users served by UAVs under given quality-of-service (QoS) requirement at each user) and minimize the number of UAVs for maintaining wireless coverage, respectively. \cite{optimal} optimized the UAVs' flying altitude to maximize the communication throughput for multiuser unicasting and multicasting communications by considering directional antennas at UAVs with adjustable beamwidth. Moreover, \cite{MM} and \cite{CZ} analyzed the average performance of quasi-stationary UAV-enabled wireless networks via the stochastic geometry theory.
For fully mobile UAVs, existing works \cite{Trajectory, Joint, J.Lyu} proposed to dynamically control their locations over time, a.k.a. trajectories, to improve the communication rates with different users at different time. For instance, \cite{Trajectory} and \cite{Joint} studied the UAV-enabled mobile relaying networks, where a UAV-mounted relay node can adaptively control its trajectory jointly with the wireless resource allocation to maximize the end-to-end communication rate from the source node to the destination node. In \cite{J.Lyu}, the authors proposed a new cyclical multiple access scheme that schedules the multiuser communication based on the UAV's trajectory, in which an interesting throughput-delay tradeoff is revealed. The authors in \cite{radio1} and \cite{radio2} used the machine learning techniques to construct radio maps for A2G wireless channels, and accordingly optimized the UAVs' trajectory for communication rate maximization. Furthermore, \cite{WPT1} and \cite{JieXuWPT} optimized the UAV's trajectory for maximizing the energy transfer efficiency towards multiple ground nodes in a UAV-enabled WPT system. \cite{LX} jointly optimized the UAV's trajectory and the transmission resource allocation to maximize the communication rates of multiple ground nodes in a UAV-enabled WPCN system, in which the energy consumption at each node cannot exceed the wireless energy harvested from the UAV. It is worth noting that both quasi-stationary and mobile UAVs have advantages and disadvantages. In particular, mobile UAVs can exploit the UAVs' fully-controllable mobility to achieve higher communication performance than quasi-stationary UAVs, but they need more sophisticated trajectory control over time with non-causal information (e.g., channel state information) required in general. By contrast, quasi-stationary UAVs only need to {\it a priori} determine the UAVs' deployment locations over a certain period of time, and thus may have much lower complexity for practical implementation. In this paper, we focus our study on the deployment optimization problem for quasi-stationary UAV-enabled wireless networks.

This paper particularly considers a scenario when multiple UAVs are employed as quasi-stationary aerial BSs to serve multiple users distributed on the ground. In practice, this network faces various technical challenges. For instance, how to provide UAVs with reliable backhaul connections to the core network is a challenging task to be tackled, as UAVs in the sky are difficult to have wireline backhauls as conventional terrestrial BSs. In the literature, although there are some prior works studying the UAV deployment optimization in UAV-enabled wireless networks (e.g., \cite{MM,CZ,2D3D1,2D3D2,optimal,radio1,radio2}), they only focused on the A2G access links between UAVs and ground users by ignoring backhaul connections. To fill such a research gap, we consider that different UAVs are connected via wireless multi-hop backhauls with the core network. Under this setup, we aim to optimize the multiple UAVs' deployment locations for network performance optimization, subject to such rate-constrained wireless backhauls. This problem, to our best knowledge, has not been investigated yet. Notice that there is one related work \cite{ws} that used UAVs to provide multi-hop wireless backhauls for connecting small ground BSs with the core network, in which the formulation of a group of UAVs is optimized for maximizing the utility of the backhaul network. Different from the prior work \cite{ws}, this paper considers both A2G access links and air-to-air (A2A) backhaul links in the UAV-enabled wireless network, for which how to jointly optimize the UAVs' deployment locations and wireless resource allocation (e.g., transmit power and bandwidth allocation) is still an open problem. This thus motivates our investigation in this work.

For the purpose of exposition, in this paper we consider that the UAVs are connected to the core network through a ground gateway node via wireless multi-hop backhauls. We suppose that the A2G access links from UAVs to users and the A2A backhaul links among UAVs are operated over orthogonal frequency bands to avoid severe co-channel interference. Due to the multi-hop connection, each UAV is subject to the so-called flow conservation constraints, i.e., the total outgoing flow from each UAV (to ground users through A2G links or to other UAVs through A2A links) should be no larger than the total incoming flow to that UAV (from the gateway node or from other UAVs). In this case, the communication rates of those ground users are fundamentally constrained by the limited rates of A2A backhaul links. Therefore, it is crucial to properly balance the rate tradeoff between the A2G access links versus the A2A backhaul links via efficient joint UAV deployment optimization and wireless resource allocation. In particular, we aim to maximize the common (or minimum) throughput among all ground users in the downlink of this network, by jointly optimizing the UAVs' deployment locations and the wireless power and bandwidth allocation in both A2A backhaul and A2G access links, subject to the individual transmit power constraints and the flow conservation constraints at the UAVs, as well as the total bandwidth constraints. However, the  common throughput maximization is a non-convex optimization problem that is generally difficult to be solved optimally. To tackle this issue, we employ the techniques of alternating optimization and successive convex programming (SCP) to optimize the UAVs' deployment locations and the wireless resource allocation in an alternating manner for obtaining a locally optimal solution. Numerical results show that the proposed design significantly improves the common throughput among all ground users as compared to other benchmark schemes.

It is worth noting that with wireless backhauls, the studied UAV-enabled wireless networks are reminiscent of the UAV-enabled relaying networks \cite{zeng2016throughput, radio1}. However, the prior works \cite{zeng2016throughput, radio1} only considered one UAV relay node with two communication hops. By contrast, this paper considers more general multi-hop connections for not only wireless backhauls among multiple UAVs but also A2G access links from UAVs to ground users. Therefore, our proposed design is more suitable for UAV-enabled cellular networks that aim to provide wireless data access for a wide range of areas with distributed ground users in e.g. emergency situations.

The remainder of this paper is organized as follows. Section \ref{sec:II} introduces the system model of the UAV-enabled wireless networks with multi-hop backhauls, and formulates the common-throughput maximization problem of our interest. Section \ref{sec:III} proposes an efficient algorithm to obtain a locally optimal solution by applying the techniques of alternating optimization and SCP. Section \ref{sec:VI} presents numerical results to validate the performance of our proposed design. Finally, Section \ref{sec:V} concludes this paper.

\begin{figure}[htb]
\centering
    \includegraphics[width=8cm]{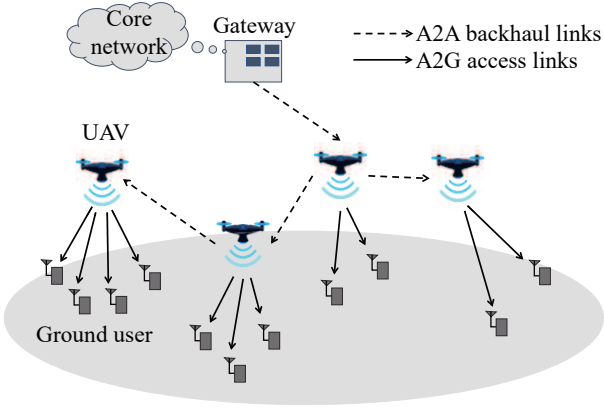}
\caption{Illustration of the UAV-enabled wireless network with multi-hop backhauls. }\label{f1}
\end{figure}

\section{System Model and Problem Formulation}\label{sec:II}

As shown in Fig. \ref{f1}, we consider a UAV-enabled wireless network, in which $M$ UAVs are deployed as aerial BSs to serve $K$ ground users, and the UAVs are connected to the core network through a gateway node on the ground via multi-hop wireless backhauls. For notational convenience, we denote the sets of UAVs and users as ${\cal M}\triangleq\left\{1, ..., M\right\}$ and ${\cal K}\triangleq\left\{1, ..., K\right\}$, respectively. We consider the 3D Cartesian coordinate system, where each user $k\!\in\!{\mathcal K}$ has a fixed horizontal coordinate $\mv{w}_k\!=\!(c_k, v_k)$ and altitude zero. Similarly as in prior works \cite{J.Lyu, zeng2016throughput, JieXuWPT}, we assume that all the UAVs are deployed at the same altitude, given by $H$, which is determined based on certain regulations on the UAVs' flying altitudes. We denote the horizontal coordinate of UAV $m\!\in\!\mathcal M$ as $\mv{u}_m\!=\!(x_m, y_m)$. Furthermore, the gateway node is fixed on the ground, with altitude zero and horizontal location $\mv{w}_0\!=\!(c_0, v_0)$. In this case, the distances from UAV $m$ to user $k$ and the gateway node are respectively denoted as
\begin{align}\label{1}
&d_{m,k}=\sqrt{H^2+\Vert\mv{u}_m-\mv{w}_k\Vert^2},\ m\in{\cal M},\ k\in{\cal K},\\
&d_{0,m}= \sqrt{H^2+\|\mv{u}_m - \mv{w}_0\|^2},\ m\in{\cal M}.
\end{align}
The distance between UAV $m$ and UAV $n$ is
\begin{align}
s_{m,n}=\sqrt{\Vert\mv{u}_m-\mv{u}_n\Vert^2}, \  m,n\in{\cal M},\  m \neq n.
\end{align}

Next, we consider the wireless transmission over the A2G access links from the UAVs to the ground users, and that over the A2A backhaul links among UAVs. For both A2G and A2A links, we suppose that the wireless channels are dominated by the LoS link, and therefore, we use the free space path loss model, as commonly adopted in the UAV communication literature \cite{zeng2016throughput, ws, Joint}. Therefore, the channel power gain from UAV $m$ to user $k$ is expressed as
\begin{align}\label{2}
\!h_{m,k}(\mv{u}_m)=\beta_0d_{m,k}^{-2}\!=\!\frac{\beta_0}{H^2+\Vert\mv{u}_m-\mv{w}_k\Vert^2},\  m\!\in\!{\cal M}, k\!\in\!{\cal K},\!
\end{align}
where $\beta_0$ is the channel power gain at the reference distance $d_0=1$ meter (m). The channel power gain from UAV $m$ to UAV $n$ is expressed as
\begin{align}\label{3}
 g_{m,n}(\mv{u}_m, \mv {u}_n)=\beta_0s_{m,n}^{-2}=\frac{\beta_0}{\Vert\mv{u}_m\!-\!\mv{u}_n\Vert^2}, \ m,n\!\in\!{\cal M},\  m\!\neq\!n.
\end{align}
Furthermore, the channel power gain from the ground gateway node to each UAV $m$ is
\begin{align}
h_{0,m}(\mv{u}_m)=\beta_0d_{0,m}^{-2}=\frac{\beta_0}{H^2+\Vert\mv{u}_m-\mv{w}_0\Vert^2},\  m\!\in\!{\cal M}.
\end{align}

Due to the strong LoS channels over both A2A and A2G links, the co-channel transmission may lead to severe interference among different users and UAVs. To overcome this issue, in this paper we consider orthogonal transmission among different A2A and A2G links to avoid any co-channel interference, in which each link is allocated with a dedicated frequency band. Let $a_{m,k}\!\ge\!0$ denote the bandwidth allocated for the A2G link from UAV $m$ to user $k$, and $p_{m,k}\!\ge\!0$ denote the transmit power of UAV $m$ over this link, $\ m\!\in\!{\cal M},\ k\!\in\!{\cal K}$. Then the achievable rate in bits-per-second (bps) of the A2G link from UAV $m\in {\cal M}$ to user $k\in {\cal K}$ is expressed as
\begin{align}\label{5}
\hat{R}_{m,k}(a_{m,k},p_{m,k},\mv{u}_m)=a_{m,k}\log_2\!\left(1+\frac{h_{m,k}(\mv{u}_m)p_{m,k}}{N_0a_{m,k}}\right)\notag \\=a_{m,k}\log_2\left(1+\frac{\gamma_0p_{m,k}}{a_{m,k}\left(H^2+\Vert\mv{u}_m-\mv{w}_k\Vert^2\right)}\right),
\end{align}
where $N_0$ denotes the noise power density at each UAV/user receiver and $\gamma_0 \triangleq  \beta_0/N_0$. In particular, we define that if $a_{m,k}=0$, then $\hat{R}_{m,k}(a_{m,k},p_{m,k},\mv{u}_m) = 0$ holds. Similarly, let $b_{m,n}\!\ge\!0$ denote the bandwidth allocated for the A2A link from UAV $m\!\in\!{\cal M}$ to UAV $n\!\in\!{\cal M}$, $m\!\neq\!n$, and $q_{m,n}$ denote the corresponding transmit power of UAV $m$ over this link. The achievable rate of the A2A link from UAV $m$ to UAV $n$ is given by
\begin{align}
\!\bar{R}_{m,n}(\!b_{m,n}, q_{m,n},&\mv{u}_m,\mv{u}_n\!)\!=\!b_{m,n}\log_2\!\left(\!1\!+\!\frac{g_{m,n}(\mv{u}_m,\mv{u}_n)q_{m,n}}{N_0b_{m,n}}\!\right)\!\!\notag
\\&=b_{m,n}\log_2\left(1+\frac{\gamma_0q_{m,n}}{b_{m,n}\Vert\mv{u}_m-\mv{u}_n\Vert^2}\right).
\end{align}
Furthermore, let $a_{0,m}\!\geq\!0$ denote the bandwidth allocated for the link from the gateway node to UAV $m$, and $p_{0,m}\!\ge\!0$ denote the transmit power of the gateway node for this link, $m\!\in\!{\cal M}$. Then the achievable rate from the gateway node to UAV $m$ is expressed as
\begin{align}
 \hat{R}_{0,m}(&a_{0,m},p_{0,m},\mv{u}_m)=a_{0,m}\log_2\left(1+\frac{h_{0,m}(\mv{u}_m)p_{0,m}}{N_0 a_{0,m}}\right)\notag
 \\&=a_{0,m}\log_2\left(1+\frac{\gamma_0p_{0,m}}{a_{0,m}\left(H^2+\Vert\mv{u}_m-\mv{w}_0\Vert^2\right)}\right).
\end{align}

Suppose that $B$ denotes the total available bandwidth for this UAV-enabled wireless network. Under the orthogonal transmission, we have the bandwidth constraint as
\begin{align}\label{4}
\sum_{m\in{\cal M}}\sum_{k\in{\cal K}}a_{m,k}+\sum_{m\in{\cal M}}\sum_{\substack{~n\in{\cal M},\\n\neq m}} b_{m,n}+\sum_{m\in{\cal M}}{a_{0,m}}\leq B.
\end{align}
Furthermore, suppose that each UAV $m\in {\cal M}$ and the gateway node are subject to the maximum transmit power $P_m\!\ge \! 0$ and $P_0\!\ge\!0$, respectively. Therefore, we have
\begin{align}
&\sum_{k\in{\cal K}}p_{m,k}+\sum_{\substack{~n\in{\cal M},\\n\neq m}}q_{m,n}\leq P_m,\ \forall m\!\in\!{\cal M},\label{7a}\\
&\sum_{m\in{\cal M}}p_{0,m}\leq P_0.\label{7b}
\end{align}

Furthermore, due to the multi-hop connections, each UAV is subject to the so-called flow conservation constraint, such that the total outgoing flows of each UAV (to users over the A2G links and to other UAVs over the A2A links) should be no larger than its total incoming flows (from the gateway and from other UAVs). It thus follows that
\begin{align}\label{6}
&\sum_{\substack{~n\in{\cal M},\\n\neq m}}\!\bar{R}_{n,m}\!(b_{n,m}, q_{n,m},\mv{u}_n,\mv{u}_m)+\!\hat{R}_{0,m}\!(a_{0,m}, p_{0,m},\mv{u}_m)\notag\\\!\geq&\!\sum_{k\in{\cal K}}\!\hat{R}_{m,k}\!(a_{m,k}, p_{m,k},\mv{u}_m\!)\!+\!\!\!\sum_{\substack{~n\in{\cal M},\\n\neq m}}\!\!\bar{R}_{m,n}\!(b_{m,n}, q_{m,n},\mv{u}_m,\mv{u}_n),\notag\\\ &~~~~~~~~~~~~~~~~~~~~~~~~~~~~~~~~~~~~~~~~~~~~~~~\forall m\!\in\!{\cal M}.
\end{align}

For notational convenience, we define $\mv{A}=\left\{a_{m,k},\ a_{0,m}, \ \forall m,k\right\}$, $\mv{B}=\left\{b_{m,n},\ \forall m,n,m\neq n\right\}$, $\mv{P}=\left\{p_{m,k},\ p_{0,m}\ \forall m,k\right\}$, $\mv{Q}=\left\{q_{m,n},\ \forall m,n,m\neq n\right\}$, and $\mv{U}=\left\{\mv{u}_m,\ \forall m\right\}$. Our objective is to maximize the common throughput among the $K$ users (i.e., $\min_{k\in K} \sum_{m\in\mathcal M} \hat R_{m,k}(a_{m,k}, p_{m,k},\mv{u}_m)$) by jointly optimizing the bandwidth allocation (i.e., \mv{A} and \mv{B}) and transmit power allocation (i.e., $\mv{P}$ and $\mv{Q}$) over both A2A links and A2G links, as well as the UAVs' deployment locations (i.e., \mv{U}), subject to the bandwidth constraint in \eqref{4}, the transmit power constraints in \eqref{7a} and \eqref{7b}, and the flow conservation constraints in \eqref{6}. Therefore, the common throughput maximization problem is formulated as
\begin{subequations}\label{9}
\begin{align}
\text{(P1)}:&\max_{\mv{A},\mv{B},\mv{P},\mv{Q},\mv{U}}
\min_{k\in{\cal K}}\sum_{m\in{\cal M}}\hat{R}_{m,k}(a_{m,k},p_{m,k},\mv{u}_m)\label{9a}\\
{\mathrm{s.t.}}\label{9b}
&~a_{m,k}\geq 0,\ {p_{m,k}\geq 0},\ \forall m\!\in\!{\cal M}, k\!\in\!{\cal K},\\\label{9c}
&~b_{m,n}\geq 0,\ {q_{m,n}\geq 0},\ \forall m,n\!\in\!{\cal M}, m\neq n,\\\label{9d}
&~a_{0,m}\geq 0,\  p_{0,m}\geq 0,\ \forall  m\!\in\!{\cal M},\\\label{9g}
&~\eqref{4}, \eqref{7a}, \eqref{7b},~\textrm{and}~\eqref{6}.
\end{align}
\end{subequations}

Note that for the common throughput maximization problem (P1), in order to deal with the flow conservation constraints in \eqref{6}, the design of the UAVs' deployment locations, and the transmit power and bandwidth allocation should optimally balance the communication rate tradeoff between the A2A backhaul links versus the A2G access links. This is a new feature that has not been investigated in prior works on UAV-enabled wireless networks. Also note that the flow conservation constraints in \eqref{6} are non-convex with respect to $\mv{A},\mv{B},\mv{P},\mv{Q}$, and $\mv{U}$, and the objective function is non-concave. Therefore, problem (P1) is a non-convex optimization problem that is generally very difficult to be solved optimally.

\section{Proposed Solution}\label{sec:III}
In this section, we propose an efficient algorithm to solve problem (P1) by using the techniques of alternating optimization and SCP.

By introducing an auxiliary variable $\eta$, we equivalently reformulate problem (P1) as
\begin{subequations}\label{10}
\begin{align}
\text{(P2)}:~&\max_{\mv{A},\mv{B},\mv{P},\mv{Q},\mv{U},\eta}~\eta\\
{\mathrm{s.t.}}
\!~\!&\sum_{m\in{\cal M}}\hat{R}_{m,k}(a_{m,k},p_{m,k},\mv{u}_m)\geq\eta,\ \forall k\!\in\!{\cal K},\label{10a}\\
&\textrm{(\ref{9b}-\ref{9g})}\label{10d}.\!
\end{align}
\end{subequations}
In the following, we deal with problem (P2) by optimizing the wireless resource allocation (i.e., the transmit power $\mv{P}$ and $\mv{Q}$, as well as the bandwidth \mv{A} and \mv{B}, and the UAVs' deployment locations \mv{U}, in an alternating manner, by considering the others to be fixed.


\subsection{Joint Transmit Power and Bandwidth Allocation Under Given UAV Deployment}\label{sec:solution2}
First, we jointly optimize the transmit power $\mv{P}$ and $\mv{Q}$ and bandwidth \mv{A} and \mv{B}, under given UAVs' deployment locations $\mv{U}$, for which the problem is expressed as
\begin{subequations}\label{11}
\begin{align}
\max_{\mv{A},\mv{B},\mv{P},\mv{Q},\eta}
~&\eta\label{11a}\\
{\mathrm{s.t.}}
~&\textrm{(\ref{10a})}~\textrm{and}~\textrm{(\ref{10d})}.
\end{align}
\end{subequations}
Problem \eqref{11} is still non-convex, which is due to the fact that the constraints in \eqref{6} are non-convex with respect to $\mv{A}$, $\mv{B}$, $\mv{P}$, and $\mv{Q}$. Therefore, problem \eqref{11} cannot be directly solved by standard convex optimization techniques. To tackle this challenge, we adopt the SCP technique to efficiently obtain a locally optimal solution. The SCP is implemented in an iterative manner by approximating problem \eqref{11} as a series of convex optimization problems. At each iteration $i\ge 1$, suppose that the local point (or the obtained solution in the previous iteration $(i\!-\!1)$) is denoted by $\mv{A}^{\left(i-1\right)}$, $\mv{B}^{\left(i-1\right)}$, $\mv{P}^{\left(i-1\right)}$, $\mv{Q}^{\left(i-1\right)}$, and $\eta^{\left(i-1\right)}$. Here, $\mv{A}^{\left(0\right)}$, $\mv{B}^{\left(0\right)}$, $\mv{P}^{\left(0\right)}$, $\mv{Q}^{\left(0\right)}$, and $\eta^{\left(0\right)}$ correspond to the initial point for iteration. Then, we need to approximate problem \eqref{11} based on such a local point as follows by approximating the non-convex constraints in  \eqref{6} into convex ones.

Notice that under given \mv{U},  $\hat{R}_{m,k}(a_{m,k},p_{m,k},\mv{u}_m)$ is jointly concave with respect to $a_{m,k}$ and $p_{m,k}$,  $\hat{R}_{0,m}(a_{0,m},p_{0,m},\mv{u}_m)$ is jointly concave with respect to $a_{0,m}$ and $p_{0,m}$, and $\bar{R}_{m,n}(b_{m,n}, q_{m,n},\mv{u}_m,\mv{u}_n)$ is jointly concave with respect to $b_{m,n}$ and $q_{m,n}$, respectively. Therefore, the constraints in \eqref{6} are non-convex, due to the fact that the right-hand-side (RHS) in \eqref{6} is jointly concave with respect to $\mv{A}$, $\mv{B}$, $\mv{P}$, and $\mv{Q}$. Fortunately, also notice that the first-order Taylor expansion of a concave function is a global over-estimator of its function value. Therefore, under any given local point $\mv{A}^{\left(i\right)}$, $\mv{B}^{\left(i\right)}$, $\mv{P}^{\left(i\right)}$, $\mv{Q}^{\left(i\right)}$, and $\eta^{\left(i\right)}$, we have
\begin{align}
~&\hat{R}_{m,k}\left(a_{m,k},p_{m,k},\mv{u}_m\right)\leq\hat{R}^{\left(i\right)-\textrm{I}}_{m,k}\left(a_{m,k},p_{m,k},\mv{u}_m\right),\\
~&\bar{R}_{m,n}\left(b_{m,n},q_{m,n},\mv{u}_m,\mv{u}_n\right)\leq\bar{R}^{\left(i\right)-\textrm{I}}_{m,n}\left(b_{m,n},q_{m,n},\mv{u}_m,\mv{u}_n\right),
\end{align}
where
\begin{align}
&\hat{R}^{\left(i\right)-\textrm{I}}_{m,k}\left(a_{m,k},p_{m,k},\mv{u}_m\right)\triangleq a^{(i)}_{m,k}\log_2\left(1+\frac{h_{m,k}(\mv{u}_m)p^{\left(i\right)}_{m,k}}{N_0a^{(i)}_{m,k}}\right)
\notag\\
&~~~~~~~~~~~~~~~~~+\left(a_{m,k}-a^{(i)}_{m,k}\right)  \log_2\left(1+\frac{h_{m,k}(\mv{u}_m)p^{(i)}_{m,k}}{N_0a^{(i)}_{m,k}}\right) \notag\\
&~~~~~~~~~~~~~~~~~~~~~-\frac{\left(a_{m,k}-a^{(i)}_{m,k}\right) h_{m,k}(\mv{u}_m)p_{m,k}^{(i)}}{\ln2\left(N_0a^{(i)}_{m,k}+h_{m,k}(\mv{u}_m)p_{m,k}^{(i)}\right)} \notag\\
&~~~~~~~~~~~~~~~~~~~~~+\frac{a^{(i)}_{m,k}h_{m,k}(\mv{u}_m)\left(p_{m,k}-p_{m,k}^{\left(i\right)}\right)}{h_{m,k}(\mv{u}_m)(a^{(i)}_{m,k}+p_{m,k}^{(i)})\ln2},\\
&\!\bar{R}^{\left(i\right)-\textrm{I}}_{m,n}\!\left(\!b_{m,n},q_{m,n},\mv{u}_m,\mv{u}_n\!\right)\!\triangleq\!
\!b^{(i)}_{m,n}\!\log_2\!\left(1\!+\!\frac{g_{m,n}(\!\mv{u}_m,\mv{u}_n\!)q^{\left(i\right)}_{m,n}}{N_0b^{(i)}_{m,n}}\right)\!
\notag\\
&~~~~~~~~~~~~~~+\left(b_{m,n}-b^{(i)}_{m,n}\right)\log_2\!\left(1\!+\!\frac{g_{m,n}(\mv{u}_m,\mv{u}_n)q^{(i)}_{m,n}}{N_0b^{(i)}_{m,n}}\right)\!\notag\\
&~~~~~~~~~~~~~~~~~~~~~-\frac{\left(b_{m,n}-b^{(i)}_{m,n}\right) g_{m,n}\!(\mv{u}_m,\mv{u}_n)\!q_{m,n}^{(i)}}{\ln2\left(N_0b^{(i)}_{m,n}+g_{m,n}(\mv{u}_m,\mv{u}_n)q_{m,n}^{(i)}\right)} \notag\\
&~~~~~~~~~~~~~~~~~~~~~+\frac{b^{(i)}_{m,n}g_{m,n}(\mv{u}_m,\mv{u}_n)\left(q_{m,n}-q_{m,n}^{\left(i\right)}\right)}{g_{m,n}(\mv{u}_m,\mv{u}_n)(b^{(i)}_{m,n}+q_{m,n}^{(i)})\ln2}.
\end{align}
By replacing $\hat{R}_{m,k}\left(a_{m,k},p_{m,k},\mv{u}_m\right)$ and $\bar{R}_{m,n}\left(b_{m,n},q_{m,n},\mv{u}_m,\mv{u}_n\right)$ as their upper bounds $\hat{R}^{\left(i\right)-\textrm{I}}_{m,k}\left(a_{m,k},p_{m,k},\mv{u}_m\right)$ and $\bar{R}^{\left(i\right)-\textrm{I}}_{m,n}\left(b_{m,n},q_{m,n},\mv{u}_m,\mv{u}_n\right)$, respectively, constraint \eqref{6} is approximated as the following convex constraints.
\begin{align}\label{17a}
&\sum_{\substack{~n\in{\cal M},\\n\neq m}}\bar{R}_{n,m}\left(b_{n,m},q_{n,m},\mv{u}_m,\mv{u}_n\right)+\hat{R}_{0,m}(a_{0,m}, p_{0,m},\mv{u}_m)\notag\\ \geq&\!\sum_{k\in{\cal K}}\!\!\hat{R}^{(i)-\textrm{I}}_{m,k}\!\left(\!a_{m,k},p_{m,k},\mv{u}_m\!\right)\!+\!\!\sum_{\substack{~n\in{\cal M},\\n\neq m}}\!\!\bar{R}^{(i)-\textrm{I}}_{m,n}\!\left(\!b_{\!m,n\!},q_{\!m,n\!},\mv{u}_m,\mv{u}_n\!\right),\notag\\
&~~~~~~~~~~~~~~~~~~~~~~~~~~~~~~~~~~~~~~~~~~~~~~~~~\forall m\in{\mathcal M}.
\end{align}

Therefore, by replacing constraints \eqref{6} as \eqref{17a}, problem \eqref{11} is approximated as the following problem:
\begin{subequations}\label{18}
\begin{align}
\max_{\mv{A},\mv{B},\mv{P},\mv{Q},\eta}
~&\eta\label{18a}\\
{\mathrm{s.t.}}
~&\textrm{(\ref{4}-\ref{7b})}, \textrm{(\ref{9b}-\ref{9d})}, \eqref{10a},~\textrm{and}~ \eqref{17a}.
\end{align}
\end{subequations}
It is evident that problem \eqref{18} is a convex optimization problem, which can thus be solved optimally by standard optimization techniques, such as the interior point method \cite{Boyd2004}.

With problem \eqref{18} solved optimally, the algorithm for solving the transmit power and bandwidth allocation problem \eqref{11} is presented as Algorithm 1 in Table I in an iterative manner. In each iteration $(i)$, we need to solve the convex optimization problem \eqref{18} under given local point $\textbf{\mv{A}}^{(i-1)}$, $\textbf{\mv{B}}^{(i-1)}$, $\textbf{\mv{P}}^{(i-1)}$, $\textbf{\mv{Q}}^{(i-1)}$, and $\eta^{(i-1)}$, for which the optimal solution is given as $\mv{A}^{\left(i\right)}, \mv{B}^{\left(i\right)}, \mv{P}^{\left(i\right)}, \mv{Q}^{\left(i\right)}$, and $\eta^{\left(i\right)}$ that will be used as the local point for the next iteration $(i+1)$. Note that the optimal value of problem \eqref{18} is an under-estimate of that of problem \eqref{11}, and therefore, the objective value of problem \eqref{11} is non-decreasing after each iteration. As the optimal value of problem \eqref{11} is upper bounded, it is evident that Algorithm 1 converges to a locally optimal solution  to problem \eqref{11}.
\begin{table}[htb]
\begin{center}
\caption{Algorithm 1 for Solving Problem \eqref{18}}  \vspace{0.1cm}
\hrule \vspace{0.3cm}
\begin{itemize}
\item[a)] {\bf Initialization}: Set the initial transmit power and bandwidth allocation as $\mv{A}^{\left(0\right)}, \mv{B}^{\left(0\right)}, \mv{P}^{\left(0\right)}$, and $\mv{Q}^{\left(0\right)}$, and set $i=1$.
\item[b)] {\bf Repeat:}
    \begin{itemize}
    \item[1)] Solve the convex optimization problem \eqref{18} under given local point $\textbf{\mv{A}}^{(i-1)}$, $\textbf{\mv{B}}^{(i-1)}$, $\textbf{\mv{P}}^{(i-1)}$, $\textbf{\mv{Q}}^{(i-1)}$, and $\eta^{(i-1)}$ via the interior point method, for which the optimal solution is given as $\mv{A}^{\left(i\right)},\mv{B}^{\left(i\right)},\mv{P}^{\left(i\right)},\mv{Q}^{\left(i\right)}$, and $\eta^{\left(i\right)}$.
    \item[2)]  $i=i+1$.
   \end{itemize}
\item[c)] {\bf Until} convergence or a maximum number of iterations has been reached.
\end{itemize}
\vspace{0.1cm} \hrule
\end{center}
\end{table}
\subsection{UAV Deployment Optimization Under Given Transmit Power and Bandwidth Allocation}\label{sec:solution4}
Next, under given transmit power and bandwidth allocation $\textbf{\mv{A}}$, $\textbf{\mv{B}}$, $ \textbf{\mv{P}}$, and $\textbf{\mv{Q}}$, we optimize the UAVs' deployment locations $\mv{U}$. The optimization problem is expressed as
\begin{subequations}\label{19}
\begin{align}
\max_{\mv{U},\eta}
~&\eta\label{19a}\\ \label{19b}
{\mathrm{s.t.}}
~&\eqref{6}~\textrm{and}~\eqref{10a}.
\end{align}
\end{subequations}
Problem \eqref{19} is also non-convex, which is due to the fact that the constraints in \eqref{6} and \eqref{10a} are non-convex. Therefore, problem \eqref{19} cannot be directly solved by standard convex optimization techniques. To tackle this challenge, we adopt the SCP technique again to efficiently obtain a locally optimal solution. At each iteration $i\ge 1$, suppose that the local point (or the obtained solution in the previous iteration $(i\!-\!1)$) is denoted by $\mv{U}^{\left(i-1\right)}$ and $\eta^{\left(i-1\right)}$, where $\mv{U}^{\left(0\right)}$ and $\eta^{\left(0\right)}$ correspond to the initial point for iteration. Then, we need to approximate problem \eqref{19} based on such a local point as follows by approximating constraints \eqref{6} and \eqref{10a} into convex constraints with respect to $\mv U$.

Notice that constraints \eqref{6} and \eqref{10a} are non-convex, due to the fact that the left-hand-side (LHS) terms in \eqref{6} and \eqref{10a} are both convex functions with respect to $\mv{U}$. As the first-order Taylor expansion of a convex function is a global under-estimator of the function value, we have
\begin{align}
&\bar{R}_{n,m}\left(b_{n,m},q_{n,m},\mv{u}_n,\mv{u}_m\right)\geq\bar{R}^{\left(i\right)-\textrm{II}}_{n,m}\left(b_{n,m},q_{n,m},\mv{u}_n,\mv{u}_m\right),\\
&\hat{R}_{0,m}\left(a_{0,m},p_{0,m},\mv{u}_m\right)\geq\hat{R}^{\left(i\right)-\textrm{II}}_{0,m}\left(a_{0,m},p_{0,m},\mv{u}_m\right),\\
&\hat{R}_{m,k}\left(a_{m,k},p_{m,k},\mv{u}_m\right)\geq\hat{R}^{\left(i\right)-\textrm{II}}_{m,k}\left(a_{m,k},p_{m,k},\mv{u}_m\right),
\end{align}
where
\begin{align}
\bar{R}^{\left(i\right)-\textrm{II}}_{n,m}&\left(b_{n,m},q_{n,m},\mv{u}_n,\mv{u}_m\right)\triangleq b_{n,m}\log_2\left(1+\frac{q_{n,m}\gamma_0}{b_{n,m}\bar{S}^{(i)}_{n,m}}\right)\notag\\
&~~~~-\frac{b_{n,m}q_{n,m}\gamma_0(\log_2e)\left(\bar{S}_{n,m}-\bar{S}_{n,m}^{(i)}\right)}{b_{n,m}\left(\bar{S}^{(i)}_{n,m}\right)^2+ q_{n,m}\gamma_0\left(\bar{S}_{n,m}^{(i)}\right)},\\
\hat{R}^{\left(i\right)-\textrm{II}}_{0,m}&\left(a_{0,m},p_{0,m},\mv{u}_m\right)\!\triangleq\! a_{0,m}\log_2\!\left(1\!+\!\frac{p_{0,m}\gamma_0}{a_{0,m}\left(H^2+\hat{S}^{(i)}_{0,m}\right)}\right)\!\notag\\
&-\!\frac{a_{0,m}p_{0,m}\gamma_0(\log_2e)\left(\hat{S}_{0,m}-\hat{S}_{0,m}^{(i)}\right)}{a_{0,m}\left(H^2+\hat{S}^{(i)}_{0,m}\right)^2+ p_{0,m}\gamma_0\!\left(H^2+\hat{S}_{0,m}^{(i)}\right)\!}\!,\\
\hat{R}^{\left(i\right)-\textrm{II}}_{m,k}&\!\left(a_{m,k},p_{m,k},\mv{u}_m\right)\!\!\triangleq\! a_{m,k}\!\log_2\!\!\left(1\!+\!\frac{p_{m,k}\gamma_0}{a_{m,k}\left(H^2\!+\!D^{(i)}_{m,k}\right)}\right)\!\notag\\&-\!\frac{a_{m,k}p_{m,k}\gamma_0(\log_2e)\left(D_{m,k}-D_{m,k}^{(i)}\right)}{a_{m,k}\left(H^2\!+\!D^{(i)}_{m,k}\right)^2\!+\! p_{m,k}\gamma_0\left(H^2\!+\!D_{m,k}^{(i)}\right)}\!.
\end{align}
Here, we define $\hat{S}_{0,m}\triangleq\|\mv{w}_0-\mv{u}_m\|^2$, $\hat{S}^{(i)}_{0,m}\triangleq\|\mv{w}_0-\mv{u}^{(i)}_m\|^2$, $\bar{S}_{n,m}\triangleq\|\mv{u}_n-\mv{u}_m\|^2$, $\bar{S}^{(i)}_{n,m}\triangleq\|\mv{u}_n-\mv{u}^{(i)}_m\|^2$, $D_{m,k}\triangleq\|\mv{u}_m-\mv{w}_k\|^2$, $D^{(i)}_{m,k}\triangleq\|\mv{u}^{(i)}_m-\mv{w}_k\|^2$, $S_{m,n}\triangleq\|\mv{u}_m-\mv{u}_n\|^2$, and $S^{(i)}_{m,n}\triangleq\|\mv{u}^{(i)}_m-\mv{u}_n\|^2$. Therefore, constraint \eqref{6} is approximated as the following convex constraint.
\begin{align}
&\!\sum_{\substack{~n\in{\cal M},\\n\neq m}}\!\bar{R}^{\left(i\right)-\textrm{II}}_{n,m}\!\left(b_{n,m},q_{n,m},\mv{u}_n,\mv{u}_m\right)\!+\hat{R}^{\left(i\right)-\textrm{II}}_{0,m}\left(\!a_{0,m},p_{0,m},\mv{u}_m\!\right)\!\notag\\ \geq&\!\!\sum_{k\in{\cal K}}\!\!\hat{R}_{m,k}\!\!\left(a_{m,k},p_{m,k},\mv{u}_m\right)\!\!+\!\sum_{\substack{~n\in{\cal M},\\n\neq m}}\!\bar{R}_{m,n}\!\!\left(b_{m,n},q_{m,n},\mv{u}_m,\mv{u}_n\right)\!.\label{39}
\end{align}
Furthermore, constraint \eqref{10a} is approximated as
\begin{align}
\sum_{m\in{\cal M}}\hat{R}^{\left(i\right)-\textrm{II}}_{m,k}\left(a_{m,k},p_{m,k},\mv{u}_m\right)\geq\eta.\label{111}
\end{align}
As a result, problem \eqref{19} is approximated as follows.
 \begin{subequations}\label{26}
\begin{align}
\max_{\mv{U},\eta}
~&\eta\label{26a}\\
{\mathrm{s.t.}}
~&\textrm{(\ref{39})}~\textrm{and}~\textrm{(\ref{111})}
\end{align}
\end{subequations}
It is evident that problem \eqref{26} is a convex optimization problem, which can thus be solved optimally by the interior point method \cite{Boyd2004}.

With problem \eqref{26} solved optimally, the algorithm for obtaining the UAVs' deployment locations in \eqref{19} is presented as Algorithm 2 in Table II. In each iteration $(i)$, we need to solve the convex optimization problem \eqref{26} under any given local point $\textbf{\mv{U}}^{(i-1)}$ and $\eta^{(i-1)}$, for which the optimal solution is given as $\mv{U}^{\left(i\right)}$ and $\eta^{\left(i\right)}$ that will be used as the local point for the next iteration $(i+1)$. Note that the optimal value of problem \eqref{26} is an under-estimate of that of problem \eqref{19}. Therefore, after each iteration, the objective value of problem \eqref{19} is non-decreasing. As a result, Algorithm 2 converges to a locally optimal solution to problem \eqref{19}.
\begin{table}[htb]
\begin{center}
\caption{Algorithm 2 for Solving Problem \eqref{26}}  \vspace{0.1cm}
\hrule \vspace{0.3cm}
\begin{itemize}
\item[a)] {\bf Initialization}: Set the initial UAVs' location as $\mv{U}^{\left(0\right)}$, and set $i=1$.
\item[b)] {\bf Repeat:}
    \begin{itemize}
    \item[1)] Solve the convex optimization problem \eqref{26} under given local point $\textbf{\mv{U}}^{(i-1)}$ and $\eta^{(i-1)}$ via the interior point method, for which the optimal solution is given as $\mv{U}^{\left(i\right)}$ and $\eta^{\left(i\right)}$.
    \item[2)]  $i=i+1$.
   \end{itemize}
\item[c)] {\bf Until} convergence or a maximum number of iterations has been reached.
\end{itemize}
\vspace{0.1cm} \hrule
\end{center}
\end{table}
\subsection{Complete Algorithm for Solving (P2)}
With Algorithms 1 and 2 at hand, we then present the complete algorithm to solve problem (P2) by solving for the wireless resource allocation (i.e., the transmit power and bandwidth allocation) and the UAV deployment, in an alternating manner. The algorithm is summarized as Algorithm 3 in Table III. Notice that for each iteration, the objective value of problem (P2) is monotonically non-decreasing, and therefore, this algorithm eventually converges to a locally optimal solution to problem (P2) (or equivalently problem (P1)).
\begin{table}[htb]
\begin{center}
\caption{Algorithm 3 for Solving Problem (P2) or (P1)}  \vspace{0.1cm}
\hrule \vspace{0.3cm}
\begin{itemize}
\item[a)] {\bf Initialization}: Set the initial UAVs' deployment locations as $\mv{U}^{\left(0\right)}$, and set $i=1$.
\item[b)] {\bf Repeat:}
\begin{itemize}
\item[1)]Use Algorithm 1 to solve problem \eqref{18} under given UAVs' deployment locations $\mv{U}^{(i-1)}$ and obtain the optimal bandwidth and power allocation as $\mv{A}^{(i)},\mv{B}^{(i)},\mv{P}^{(i)}$, and $\mv{Q}^{(i)}$.
\item[2)]
Use Algorithm 2 to solve problem \eqref{26} under given transmit power and bandwidth allocation $\mv{A}^{(i)}, \mv{B}^{(i)}, \mv{P}^{(i)}$, and $\mv{Q}^{(i)}$,  and obtain the optimal UAVs' deployment locations $\mv{U}^{(i)}$.
\item[3)] $i=i+1$.
\end{itemize}
\item[c)] {\bf Until} convergence or a maximum number of iterations has been reached.
\end{itemize}
\vspace{0.1cm} \hrule
\end{center}
\end{table}

\section{Numerical Results}\label{sec:VI}
In this section, we present numerical results to validate the performance of our proposed design. In the simulation, we set the system bandwidth as $B=10$ MHz, the carrier frequency as $f_c=5$ GHz, and the noise power density as $-169$ dBm/Hz. Accordingly, the channel power gain at the reference distance $d_0=1$ m is $\beta_0=\frac{c}{{\left(4\pi f_{c}\right)}^2}$, with $c=3\times 10^8$ m/s denoting the light speed. We set the flying altitude of UAVs as $H=100$ m. Furthermore, the total transmit power at each UAV and the gateway node are set as $P_{m}=P_0=P$, $\forall m\in M$. Under this setup, we compare the performance of our proposed design versus the following two benchmark schemes.
\begin{itemize}
\item {\it Joint transmit power and bandwidth allocation}: This scheme jointly optimizes the transmit power $\mv{P}$ and $\mv{Q}$, and the bandwidth $\mv{A}$ and $\mv{B}$ under fixed UAVs' locations. This corresponds to solving problem \eqref{18} via Algorithm 1. In the simulation, the UAV are set to be evenly distributed within the area of our interest, which are also used as the initial point for our proposed Algorithm 3 to solve problem (P2), as will be specified later.
\item {\it UAVs' deployment optimization}: This scheme only optimizes the UAVs' deployment locations (i.e., \mv{U}) under fixed bandwidth allocation $\mv{A}$ and $\mv{B}$ and transmit power allocation $\mv{P}$ and $\mv{Q}$. This corresponds to solve problem (23) based on Algorithm 2. In the simulation, the bandwidth and transmit power are equally allocated among different communication links.

\end{itemize}

\subsection{One-Dimensional (1D) User Distributions Case}
\begin{figure}[h]
  \centering
  \includegraphics[width=7.5cm]{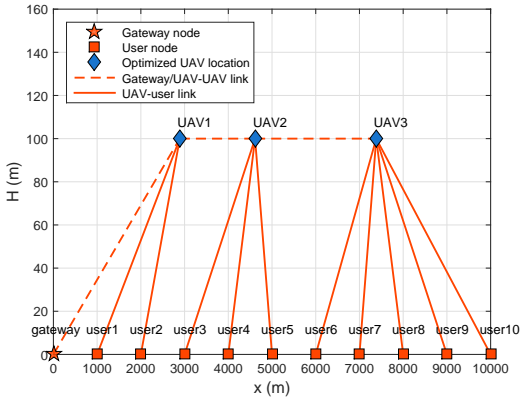}\\
  \caption{Simulation setup: the case with 1D user distributions.}\label{f1D}

\end{figure}

First, we consider the case with 1D user distributions in Fig.~\ref{f1D}, where there are $M = 3$ UAVs serving $K = 10$ users distributed in a 1D space, and the transmit power of each node is set as $P=30$ dBm. For our proposed Algorithm 3, the initial deployment locations of the three UAVs are set as (2500m, 100m), (5000m, 100m), and (7500m, 100m), respectively. These locations are also used for the benchmark scheme with joint transmit power and bandwidth allocation. Fig.~\ref{f1D} shows the optimized UAVs' deployment locations under our proposed design. It is observed that the A2A backhaul links are connected in a multi-hop nature from the gateway to UAV $1$, to UAV $2$, and finally to UAV $3$. It is also observed that when the UAV is closer to the gateway, it serves less users at the A2G access links. More specifically, UAVs $1$, $2$, and $3$ are observed to serve $2$, $3$, and $5$ users, respectively. This is due to the fact that when the UAV is closer to the gateway, it needs to split more wireless power and bandwidth resources to support the A2A links, in order to better balance the rate tradeoff between A2A backhaul and A2G access links.

\begin{figure}[htb]
  \centering
  \includegraphics[width=7.5cm]{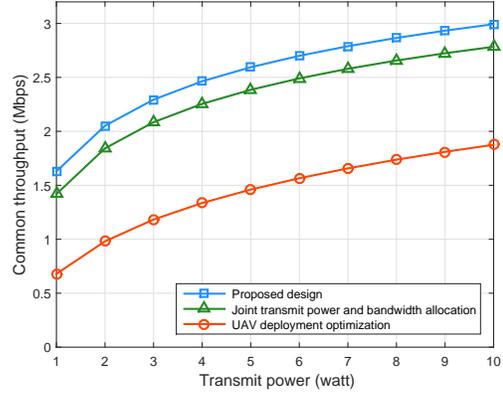}\\
  \caption{Common throughput of ground users versus transmit power $P$ with 1D user distributions.}\label{p1D}
\end{figure}

Fig. \ref{p1D} shows the common throughput of the ground users versus the transmit power $P$ in the 1D case. It is observed that the proposed design outperforms the other two benchmark schemes. This shows the benefit of our design with joint transmit power, bandwidth, and UAV deployment optimization. It is also observed that the scheme with joint transmit power and bandwidth allocation achieves superior performance to the scheme with UAV deployment optimization.

\subsection{2D User Distributions Case}
\begin{figure}[htb]
  \centering
  \includegraphics[width=7.5cm]{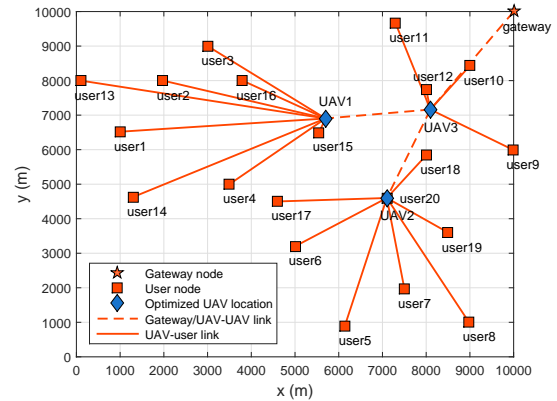}\\
  \caption{Simulation setup: the case with 2D user distributions.}\label{f2}

\end{figure}

Next, we consider the case with 2D user distributions, where there are $M=3$ UAVs and $K=20$ users within a square area with $10$ km $\times$ $10$ km. The horizontal locations of the users and the ground gateway node are shown in Fig. \ref{f2}. For our proposed Algorithm 3, the initial deployment locations of the three UAVs are set as (3333m, 6666m, 100m), (6666m, 3333m, 100m), and (6666m, 6666m, 100m), respectively.

Fig.~\ref{f2} shows the optimized UAVs' deployment locations via our proposed design, in which the transmit power is set as $P = 30$ dBm. It is observed that UAV $1$ is deployed close to users $1-4$ and users $13-16$ for serving them, UAV $2$ serves users $5-8$ and users $17-20$, and UAV $3$ serves users $9-12$, respectively. It is also observed that UAV $1$ and UAV $2$ are connected to the gateway node via relaying through UAV $3$. The optimized UAVs' deployment locations and corresponding wireless resource allocation efficiently balance the rate tradeoff between A2G access and A2A backhaul links.

\begin{figure}[htb]
  \centering
  \includegraphics[width=7.5cm]{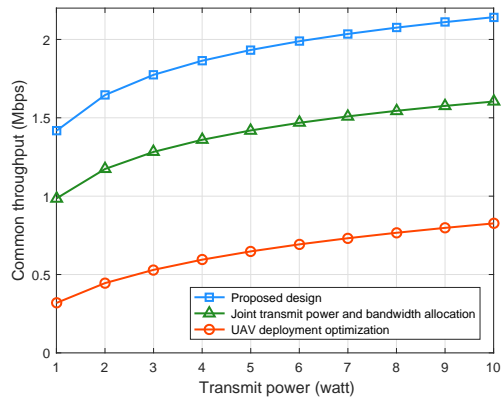}\\
  \caption{Common throughput of ground users versus transmit power $P$ with 2D user distributions.}\label{f4}
\end{figure}

Fig. \ref{f4} shows the common throughput of the ground users versus the transmit power $P$ for the 2D case. Similar observations are made here as those in Fig. \ref{p1D} for the 1D user distribution case. Furthermore, it is observed that our proposed design achieves more significant performance gain over the benchmark schemes in 2D case, as compared with that in the 1D case as shown in Fig.~\ref{p1D}. This is due to the fact that in the 2D case, we can exploit more degrees of freedom in the UAVs' deployment locations optimization, thus leading to better common throughput performance.


\section{Conclusion}\label{sec:V}
In this paper, we studied a UAV-enabled wireless network with wireless multi-hop backhauls. By considering the flow conservation constraint at UAVs, we jointly optimized the transmit power and bandwidth allocation, as well as the UAVs' deployment locations, in order to maximize the common throughput among all ground users. By applying the techniques of alternating optimization and SCP, we present an efficient algorithm to solve this problem, which is guaranteed to converge to at least a locally optimal solution. Numerical results demonstrate that the proposed design significantly improves the common throughput among all ground users as compared to other benchmark schemes. It is our hope that this paper can provide new insights on the design of UAV-enabled wireless networks, in which the consideration of rate-constrained backhauls are crucial for the performance optimization.

\begin{small}

\end{small}
\end{document}